\documentclass[]{elsart}
\usepackage{graphicx}
\newcounter{bla}
\newenvironment{refnummer}{%
\list{[\arabic{bla}]}%
{\usecounter{bla}%
 \setlength{\itemindent}{0pt}%
 \setlength{\topsep}{0pt}%
 \setlength{\itemsep}{0pt}%
 \setlength{\labelsep}{2pt}%
 \setlength{\listparindent}{0pt}%
 \settowidth{\labelwidth}{[9]}%
 \setlength{\leftmargin}{\labelwidth}%
 \addtolength{\leftmargin}{\labelsep}%
 \setlength{\rightmargin}{0pt}}}
 {\endlist}

\begin{document}
\begin{frontmatter}

\title{Pretty Fast Analysis: An embarrassingly parallel algorithm for biological simulation analysis}

\author[a]{David N. LeBard\thanksref{author}},

\thanks[author]{Corresponding author}

\address[a]{Center for Biological Physics, Arizona State University, 
PO Box 871604, Tempe, AZ 85287-1604}

\begin{abstract}
A parallel code has been written in FORTRAN90, C, and MPI for the analysis of biological simulation data.  Using a master/slave algorithm, the software operates on AMBER generated trajectory data using either UNIX or MPI file IO, and it supports up to 15 simultaneous function calls.  This software has been performance tested on the Ranger Supercomputer on trajectory data of an aqueous bacterial reaction center micelle.  Although the parallel reading is poor, the analysis algorithm itself shows embarrassingly parallel speedup up to 1024 compute nodes.  At this CPU count the overall scaling of the software compares well NAMD's best reported speedup, and outperforms AMBER's best known scaling by a factor of 3, while using only a small number of function calls and a short trajectory length. 

\begin{flushleft}
\end{flushleft}


\end{abstract}

\end{frontmatter}


{\bf PROGRAM SUMMARY}

\begin{small}
\noindent
{\em Program Title:} Pretty Fast Analysis                    \\
{\em Journal Reference:}                                      \\
{\em Catalogue identifier:}                                   \\
{\em Licensing provisions:} Standard CPC license, http://cpc.cs.qub.ac.uk/license/license.html\\
{\em Programming language:} FORTRAN90/C/MPI                                  \\
{\em Computer:} Platform independent                                         \\
{\em Operating system:} OS independent                         \\
{\em RAM:} Typically a maximum of 250 MB on the nodes and 2 GB on the master     \\
{\em Number of processors used:} 1 to 1024.            \\
{\em Supplementary material:}                                 
{\em Classification: 3}                                         \\
{\em Nature of problem:} Currently, several parallel biomolecular simulation packages are capable of generating literally terabytes of trajectory data.  However, there has not been a corresponding push to create parallel biomolecular simulation analyzer that can handle large trajectory sets in a reasonable time.\\
{\em Solution method:} This software calculates both structural and energetic properties of biomolecules in parallel using a simple master/slave methodology.  The program itself is keyword driven, allowing up to 15 different structural and energetic calculations to be performed at once.  Trajectories can be read in either binary or ASCII formats, using either typical Unix or MPI IO.  This software has been through several revisions, and in its current state it operates well in either a small linux cluster or supercomputer environment.  In the past, the code has been used to calculate the thermodynamic and kinetic properties of aqueous tryptophan amino acid [1], plastocyanin [1,2] and the bacterial reaction center [3], where the latter two are canonical proteins studied within the umbrella of protein electron transfer. \\
{\em Restrictions:} In the current form, the code only analyzes AMBER trajectory data from the binpos, mdcrd, or modified binpos filetypes.  Also, the code has only been tested with
Portland Group compilers and therefore mixing other FORTRAN/C compilers is not recommended.\\
{\em Additional comments:} The AMBER topology file reader subroutine (readprm), was taken from readprm.f from the AMBER8 distribution and has been used with permission from David Case and the AMBER development team.\\
{\em Running time: } Depends on the size of the biomolecule, the number of the surrounding solvent  molecules, the nature of the calculations, as well as the number of frames in the simulation trajectory.  Generally, a trajectory of $500,000$ frames can be analyzed for tens of energetic and structural properties in a matter of hours.\\
{\em References:}
\begin{refnummer}
\item D. N. LeBard, D. V. Matyushov, J. Phys. Chem. B. 112 (2008) 5218         
\item D. N. LeBard, D. V. Matyushov, J. Chem. Phys. 128 (2008) 155106         
\item D. N. LeBard, V. Kapko, D. V. Matyushov, J. Phys. Chem. B. (2008) ASAP 
\end{refnummer}
\end{small}

\newpage


\hspace{1pc}
{\bf LONG WRITE-UP}

\section{Introduction}
\paragraph*{}
For more than 30 years, advancements in biological molecular dynamics (MD) simulations have pushed the limits of computation in terms of parallel scalability \cite{namd99,amber8,charmm83} and algorithm optimization\cite{namd99,pme95}.  Biological systems under study will generally consist of $N_{sys} \sim 10^4-10^6$ atoms when run with medium to large biomolecular polymers in explicit or implicit model solvents.  In recent years, data have shown that in order to capture the long time dynamical motions of biological systems \cite{pcjpcb08,rcjpcb08}, production runs on the order of nanoseconds to milliseconds are required, stretching the MD performance requirements even further.  The simulation algorithms themeselves, though very efficient, are highly specialized to achieve maximum performance on the largest number of processors\cite{namd99} to generate a trajectory of configuration data by numerically solving equations of motion.  The basic strategy of every MD software is to parallelize the force calculations by using a spacial or particle-based decomposition, and in some cases using dynamic load balancing to achieve highest level of scalability \cite{namd99} and consequently, the quickest trajectory generation.

While these advancements have allowed studies of long time dynamics of biomolecular systems with atomistic detail, similar attention has not been paid to the analysis of such large data sets.  Trajectories produced from biological simulations contain configuration data of all $N_{sys}$ atoms in the system for a given number of steps, $N_{steps}$, around $10^5-10^6$ for runs in the nanosecond range for a moderate saving frequency.  To analyze such trajectories, pairwise interactions between subsets ($N_{sys} = N_1 + N_2 + ... $) of these atoms are calculated at each timestep such that the total number of calculations, and therefore the analysis time, are on the order of 

\begin{equation}
  \label{eq:1}
  \tau _{ana} \propto O(\gamma N_1N_2N_{steps})
\end{equation}

Here, $\tau _{ana}$ is the total analysis time and $\gamma$ is the number of function calls required for the analysis.  In principle, $N_1$ and $N_2$ differ for each function call during the analysis procedure.  If the number of needed pairwise calculations are large enough, parallelizing the analysis at the $N_{steps}$ or $N_{sys}$ level would certainly lead to an strongly scaling algorithm where the analysis time would be reduced to 

\begin{equation}
  \label{eq:2}
  \tau _{ana}(N_{CPUs}) = \tau _{ana}/N_{CPUs}
\end{equation},

where $N_{CPUs}$ refers to the number of CPUs available for the calculation.  

To handle such large number of calculations, a parallel data analysis algorithm has been developed to perform a variety of structural and energetic calculations to decrease the time to solution of various biophysical problems proportional to the number of compute nodes invested.  This algorithm, built into the Pretty Fast Analysis (PFA) software, assumes that the total analysis time follows the simple relation:

\begin{equation}
  \label{eq:3}
  \tau _{tot}(N_{CPUs}) = \tau _{ana}(N_{CPUs}) + \tau _{read} + \tau _{comm}
\end{equation}

In Eq. \ref{eq:3}, $\tau _{tot}$ is the total software time, $\tau _{read}$ is the time required for trajectory reading, and $\tau _{comm}$ is the time for any associated time for communication between processors.  One would expect that $\tau _{tot} > \tau _{ana} > \tau _{comm}$.  By parallelizing trajectory reading analogous to the data analysis (Eq. \ref{eq:2}), and by minimizing communication costs such that $\tau _{comm} \ll \tau _{read}$, one can see that the speedup (the ratio $\tau _{tot}(1CPU)/\tau _{tot}(N_{CPUs})$) will be equal to $N_{CPUs}$.  The goal of this paper will be to demonstrate this relationship holds experimentally true using up to thousands of compute nodes for a minimum number of function calls and trajectory steps.

After a brief theoretical overview of the PFA analysis applied to electron transfer theory in Section 2, the software itself will be overviewed in section 3.  The software's parallel performance will be shown in section 4, and we will reveal the final conclusions in section 5.  For more detailed instructions on running the software, the reader is referred to the PFA website.

\section{Theoretical Background}
\paragraph*{}
This code was born out of necessity to analyze large amounts of simulation data on several photosynthesis proteins.  Therefore the type of calculations given here is merely an example from electron transfer theory to provide a real-life example of the extent to which the analysis may be performed in parallel.  This section is not meant to be an introduction to electron transfer theory, for that the reader should see the review in Ref. \cite{matyushov07}, but rather as an introduction to the calculations used as the test of the PFA algorithm.

In electron transfer theory, the first and second cumulants of the vertical energy gap, $\Delta E$, are needed to calculate the kinetic and thermodynamic parameters activating the electron transfer process \cite{pcjpcb08,pcjcp08,rcjpcb08}.  The energy gap itself is defined as:

\begin{equation}
  \label{eq:4}
  \langle \Delta E \rangle = \Delta E^C + \Delta E^I + \Delta E^{gas}
\end{equation}

The vertical energy gap in Eq. \ref{eq:4} describes the change in electrostatic energy at the enzymatic site during the charge transfer process and has three components, two of which, $\Delta E^C$ and $\Delta E^I$, can be readily calculated from the simulation data.  The first term in Eq. \ref{eq:4} represents the Coulomb component of the vertical energy gap, and can be written as

\begin{equation}
  \label{eq:5}
  \Delta E^C = \sum_j \Delta q_j \phi_j .
\end{equation}

In this equation, the index $j$ runs over all atoms from the redox site that can change electronic state, and $\phi$ represents the potential of the surrounding biological solvent.  Free energy surfaces of oxidation-reduction processes can also be quite affected by the high polarizability of certain cofactors found in electron transfer proteins\cite{matyushov07,rcjpcb08,small03}.  In fact, this term can be calculated from a standard force field simulation \cite{rcjpcb08} using an energetic term accounting for the induction forces of the shifting charge.  Since atomic polarizabilities of the reaction center can be tabulated, one needs to calculate the induction term of the vertical energy gap, $\Delta E^I$, given in the following form:

\begin{equation}
  \label{eq:6}
  \Delta E^I = - \left\langle \sum_j (\alpha _j/2) [E_{02} ^2(r_j) - E_{01} ^2(r_j)] \right\rangle.
\end{equation}

In Eq. \ref{eq:6} the sum runs over all atoms in the redox site, and $E_{0X}$ refers to the electric field of enzymatic center the protein in the $X$ state.  The electric field is evaluated at a distance $r_j$ from the active site in the solvent, and $\alpha _j$ refers to the atomic polarizabilities.  

It has recently been suggested that the dynamics of electron transfer follows the fluctuations of the donor-acceptor distances between electron transferring cofactors \cite{chaudhury07}.  To test this theory in simulations, the separation distance between acceptor and donor cofactors, $\vec{r}_{DA}$, is required.  This is simply a vector subtraction in the center of mass frame is given by 

\begin{equation}
  \label{eq:7}
  \vec{r}_{DA} = \vec{R}_{Donor} -  \vec{R}_{Acceptor}
\end{equation}

Here, $R_{Donor}$ and $R_{Acceptor}$ are the mass weighted vectors of the donor and acceptor cofactors, respectively.  The same donor-acceptor atoms vary their charges during the redox process, and therefore it is quite useful to monitor the charge transfer dipole moment

\begin{equation}
  \label{eq:8}
  \Delta \vec{m}_{DA} = \sum_{j \in DA} \Delta q_j \vec{r}_j
\end{equation}

In Eq. \ref{eq:8}, $\Delta q_j$ are the difference charges and the sum runs over all atoms in either an electron donor or acceptor cofactor.  These equations provide a simple, yet practical example of the efficiency of a highly parallelizable analysis algorithm.  

\section{Overview of the software}
\begin{figure}  
  \centering
  \includegraphics*[width=8cm]{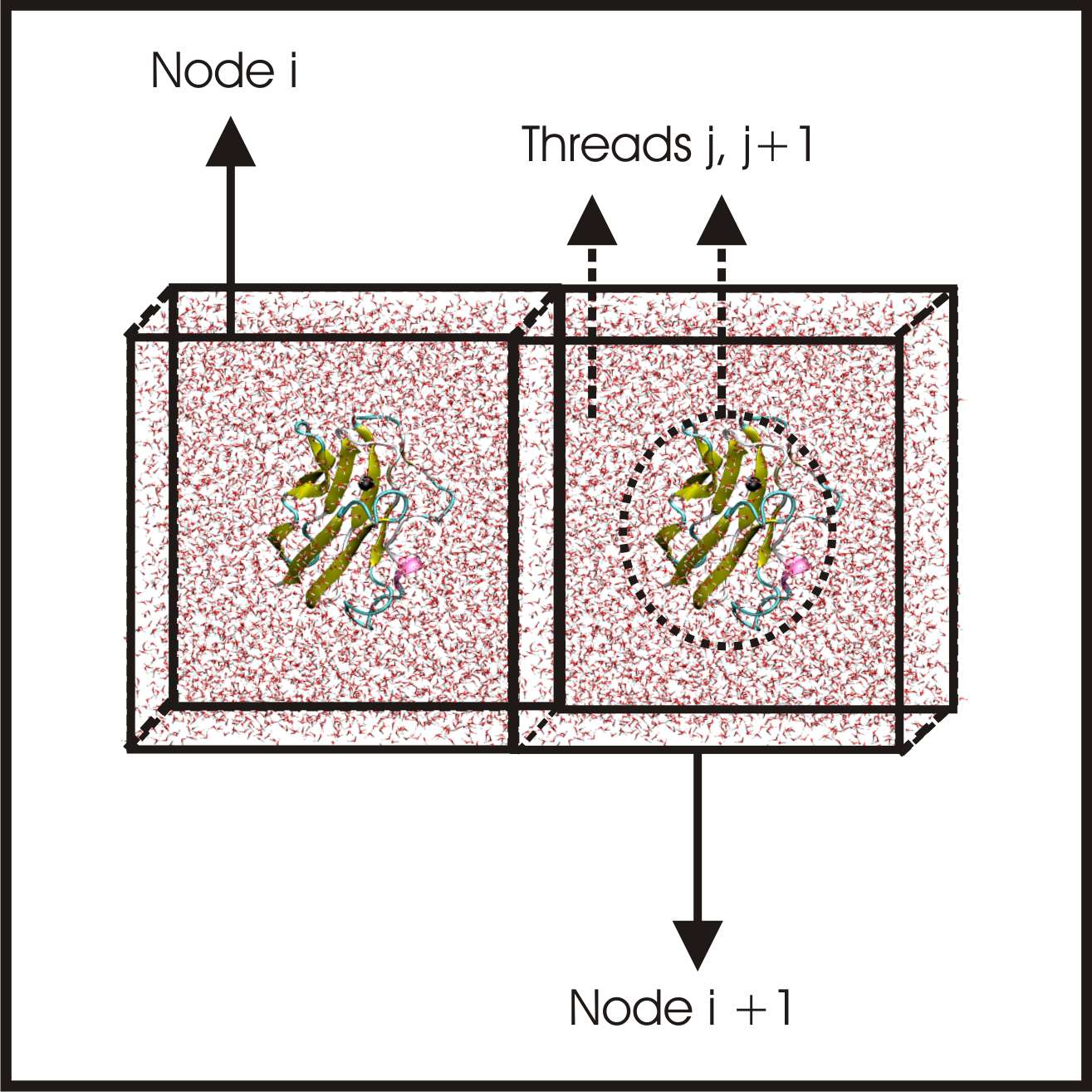}
  \caption{Schematic of analysis data parallelization. The current version of the algorithm scales strongly and is illustrated with the thick arrows.  This method distributes configuration data across the nodes, and each node calculates the all energetics and structural properties of the system. Shown with the dashed arrows, one can see how the data can be further parallelized using threads.  In this protocol, properly initialized shared memory threads can calculate pairwise calculations without any send cost, and therefore reduce the analysis time to $\tau_{ana}(N_{CPUs}) /N_{Threads}$, while concurrently reducing the number of communications by $1/N_{Theads}$ by splitting the computations into those of the solvent (thread j) and the protein (thread j+1)}
\label{fig:para}
\end{figure}

\paragraph*{}
The data analysis software has proven to be extremely fast, even over the course of its two year development limetime due to the simple data splitting procedure (Figure \ref{fig:para}).  As this figure suggests, the data is distributed such that each node has the entire configuration data for an individual frame of the trajectory.  This scheme ensures a minimum communication overhead, which would be incurred if each CPU had less than $N_{sys}$, at the cost of storing a dataset on the order of O($N_{sys}$).

\begin{figure}
  \centering
  \includegraphics*[width=8cm]{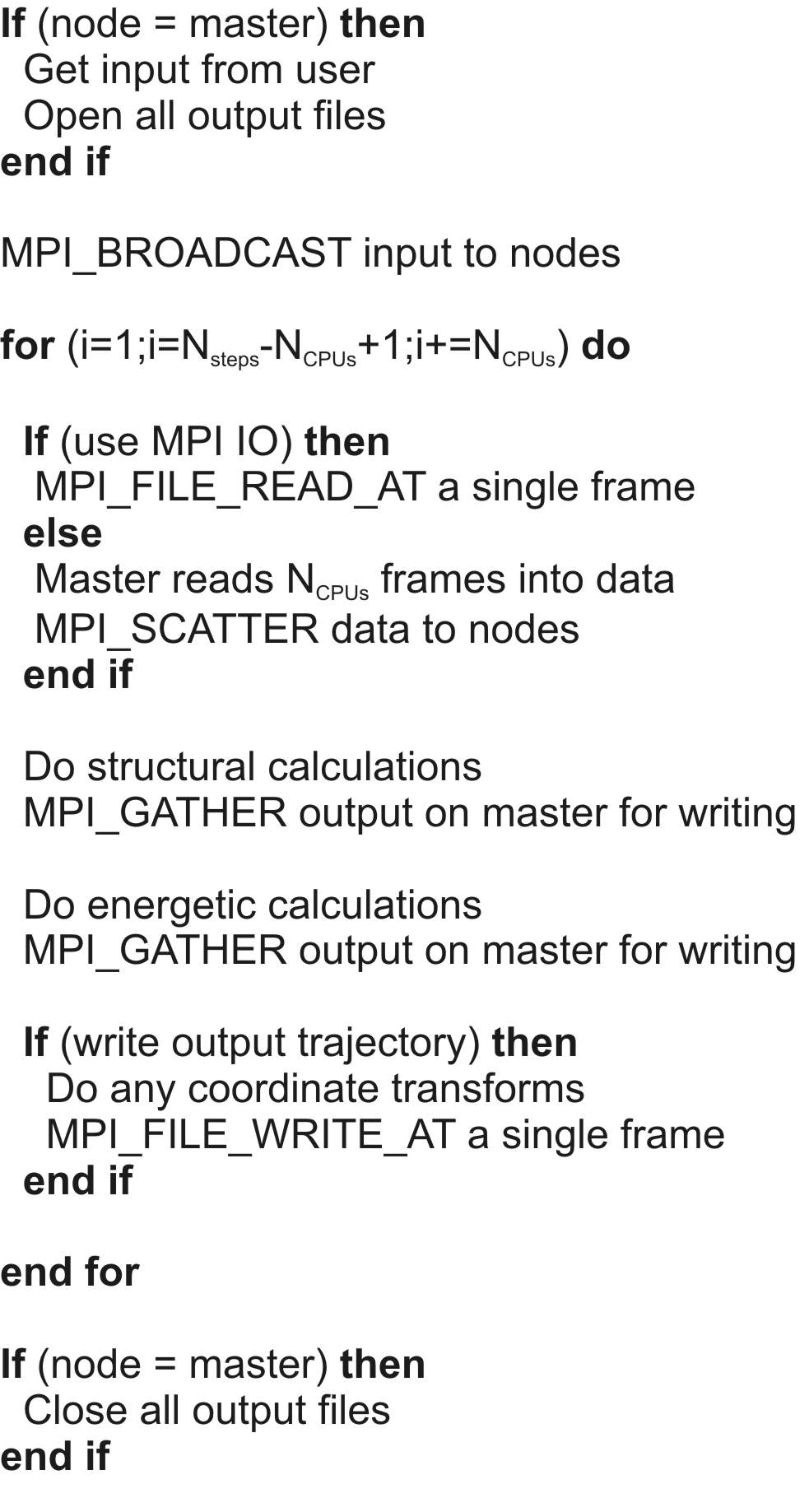}
  \caption{Pseudocode of the PFA algorithm.  To invoke threads, a decision to use them is taken at input and threads are initialized before the main loop.  The threads themselves will need to decide a master thread for each node for sending and receiving the analysis data.}
  \label{fig:para-algo}
\end{figure}

A the time of writing, the code consists of $9,950$ lines spread over 9 FORTRAN files, and a single file of C code used for .gz file reading.  After compiling the source using the make command, an executable named pfa.e is generated and can be controlled at runtime using a simple keyword driven interface.  The code itself is run like,

\begin{verbatim}
mpirun -np <number of cpus> pfa.e < pfa.in >& pfa.out 
\end{verbatim}

In its current form, the software can perform up to 15 analysis functions at once including vertical energy gap calculations, vector monitoring, dipole moment calculations, water-shell calculations, mean-squared displacement calculations, as well as AMBER trajectory conversion using MPI IO.  The code is capable of reading trajectory data from AMBER's  binpos, modified binpos, ASCII mdcrd, or gzipped mdcrd formats with either standard UNIX or MPI IO.  It should be noted that Gustafson's Law \cite{gustafson88} allows PFA to scale most strongly when all calculations are made in a single run, which could be a considerable efficiency burst for extremely large systems or those with many atoms in subsets $N_1$ and $N_2$.  In the example shown below, the code can be driven to embarrassingly parallel speedup by using even a modest number of function calls (4) for a bacterial reaction center micelle in an explicit TIP3P solvent \cite{tip3p83}.  For the details of the simulation protocol, the reader is referred to previous work on this system\cite{rcjpcb08}.

\section{Performance results}
\paragraph*{}
The PFA software has been well tested on several clusters ranging in speed from a local Opteron cluster with a gigabit ethernet network, to the Saguaro Supercomputer at Arizona State University running and infiniband network, to the the newly constructed Ranger Supercomputer at TACC.  Although this paper focuses on the timing analysis using MPI-IO on the Ranger Supercomputer, similar results have been seen in all three test environments, using UNIX and MPI IO, and the details of such comparisons can be found in the code's manual.

\begin{table}
  \centering
  \caption{{\label{tab:1}} Timing data for the PFA test on the Ranger supercomputer. All timing data is given in seconds, and has been measured using the WTIME function from the MPI library.}
  \begin{tabular}{cccc}
$N_{CPUs}$ & $\tau _{read}$ & $\tau _{ana}$ & $\tau _{tot}$ \\
\hline
\hline
1 &  236.6 &  148909.1 &  149145.7 \\
16 &  246.7 &  9371.0 &  9617.7 \\
32 &  92.4 &  4683.4 &  4775.8 \\
64 &  66.3 &  2349.8 &  2416.1 \\
128 &  61.2 &  1181.9 &  1243.1 \\
256 &  67.1 &  590.2 &  657.3 \\
512 &  72.4 &  294.3 &  366.7 \\
1024 &  74.5 &  147.0 &  221.4 \\

\hline
  \end{tabular}
\end{table}

To test the scalability of the software, a short 12,288 step (245.76 ps) trajectory from a recent simulation of aqueous micelluar bacterial reaction center (BRC) protein\cite{rcjpcb08} was used.  The very short trajectory length was determined to be the maximum trajectory length a single CPU could run the 4 functions in a 48 hour period, and therefore was the longest trajectory that could be used.  This test provided enough computations to illustrate the high efficiency of the algorithm at thousands of CPUs.  The test analysis consisted of the difference dipole moment of the redox site and the center of mass distance between two electron transfer cofactors, as well as the Coulomb and Induction contributions to the vertical energy gap as given in equations \ref{eq:4}--\ref{eq:8}.  The results of the calculations have also been covered elsewhere\cite{rcjpcb08}, so only the scalability of the analysis itself will be discussed here.  All timing data for the reading time, analysis time, and full software analysis time is provided in Table \ref{tab:1}.

\begin{figure}
  \centering
  \includegraphics*[width=10cm]{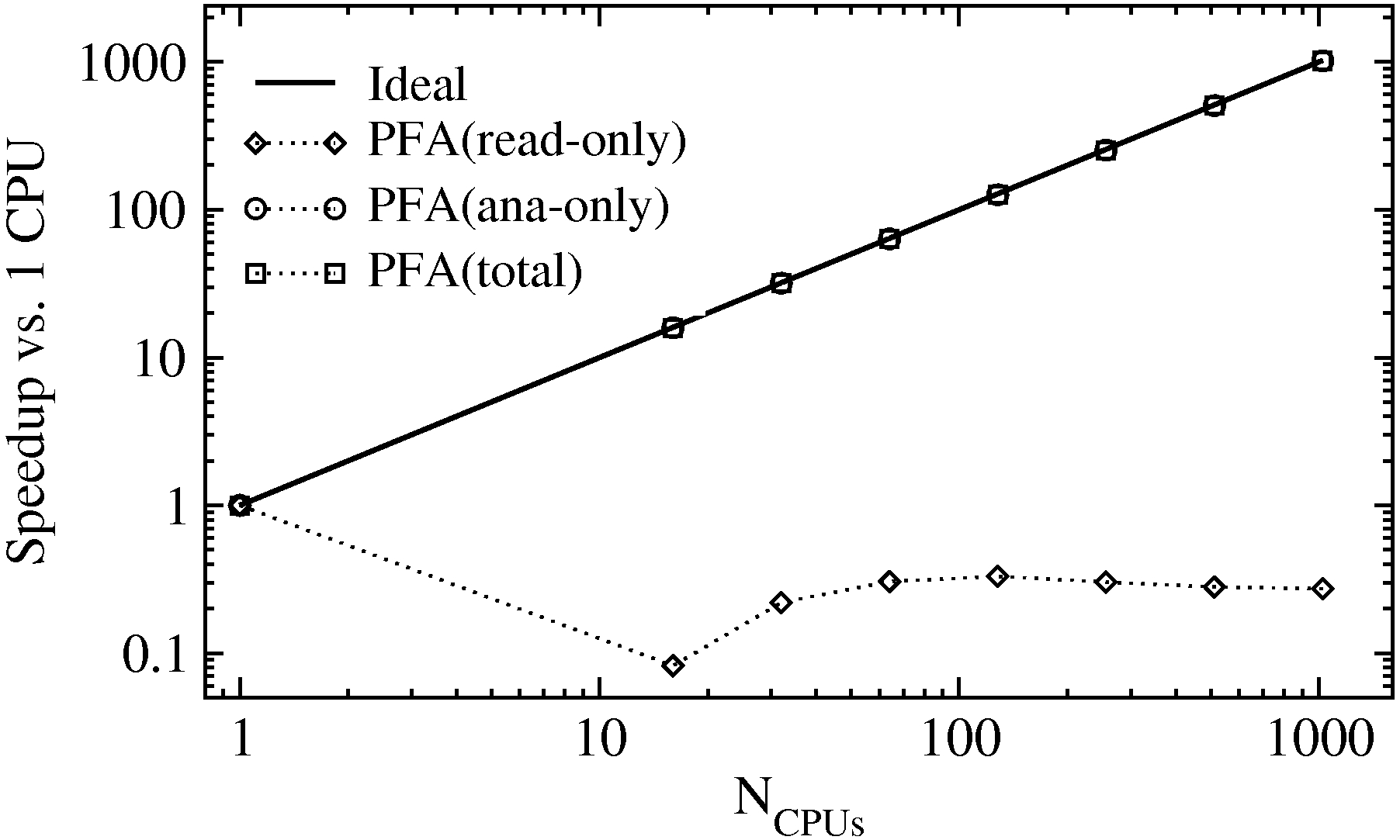}
  \caption{Breakdown of the timing of PFA on the Ranger Supercomputer.  These timing analysis have been measured using MPI's WTIME function.}
  \label{fig:pfa-speed}
\end{figure}

The speedup data for this test on Ranger is shown in Fig. \ref{fig:pfa-speed}.  From the splitting of the performance data into its three principle components, one will notice that the speedup of the reading in parallel is quite poor (diamonds in Fig. \ref{fig:pfa-speed}).  Even though the reading time generally decreased with increasing CPUs, the reading time cannot be decreased to less than around 60s for this particular test.  However, the reading times are somewhat linearly decreasing in the range of $N_{CPUs}=16-128$ (see Table \ref{tab:1}), suggesting that the shortness of the trajectory skews the reading parallelizabilty to appear less scalable.  In other words, longer trajectories will lead to an increase in speedup because the time of reading the total trajectory is no longer on the order of the MPI IO initialization time, or more importantly, on the order of the analysis time.  

In contrast to the poor scalability seen for reading, the analysis algorithm shows \textit{embarrassingly parallel} scalability (circles in Fig. \ref{fig:pfa-speed}).  Over the entire test range, from $1-1024$ CPUs, the speedup of the analysis algorithm itself traces the ideal linear case.  Ideal speedup is achieved for the analysis because of the large system size (ca. 50,000 atoms), the large number of pairwise calculations due to the invoked function calls, and the trajectory length is long enough to ensure reasonable performance on a large number of CPUs.  Therefore, when analyzing a nanosecond or longer trajectory  $\tau _{read} \ll \tau _{ana}$, and parallel reading no longer affects the scalability.  However achieving this kind of speedup is not unusual, but instead should be considered the minimum performance one can achieve with PFA.  In fact, several calculations involving search routines over the first solvation shell of the protein are quite time consuming, and one can expect embarrassingly parallel scalability over larger number of processors with small systems and short ($<$ 1 ns) trajectories.

\begin{figure}
  \centering
  \includegraphics*[width=10cm]{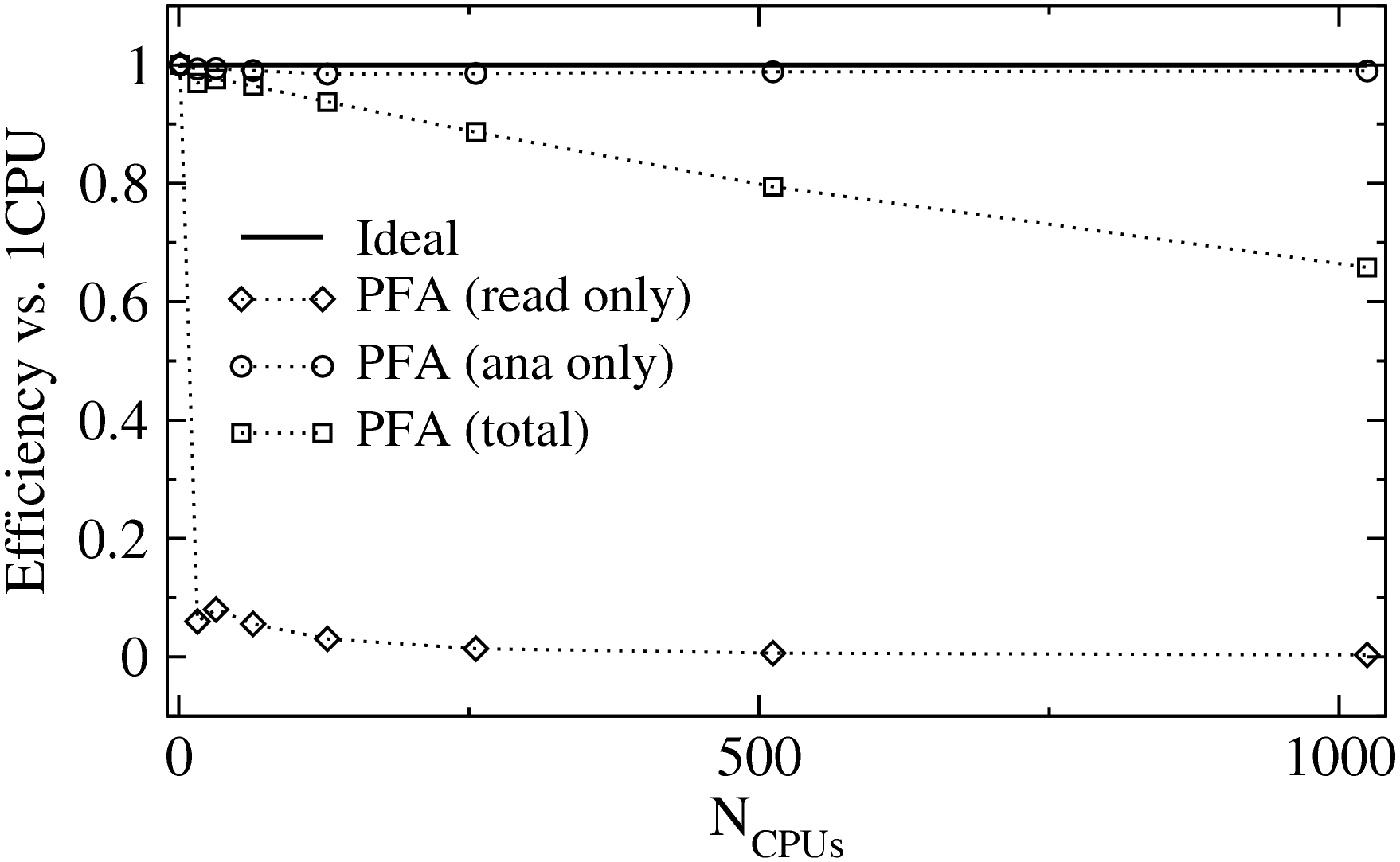}
  \caption{Breakdown of the efficiency of PFA on the Ranger Supercomputer.  Although the reading is considerably inefficient, the analysis algorithm itself is so efficient that the entire program is more than 98 \% efficient over thousands of processors.}
  \label{fig:pfa-eff}
\end{figure}

The poor scaling of the parallel reading degrades the overall performance of the total program (squares in Fig. \ref{fig:pfa-speed}) considerably.  In the region of linear scaling of the reading times, generally less than a few hundred processors, the total program scales linearly as well.  However, once the reading time cannot be lowered with a larger number of processors and it becomes roughly $1/3$ of the total time of the total program, the program cannot scale due to the limitations of the IO interface.

As one would expect the parallel performance of the algorithm creates an environment for high efficiency, which can defined as $Speedup(N_{CPUs})/N_{CPUs}$ and is given for the Ranger test in Fig. \ref{fig:pfa-eff}.  Despite the fact that the reading code slows to maximum inefficiency after a small number of CPUs (diamonds), the analysis algorithm is roughly 99 \% efficient over 1024 CPUs.  Such efficiencies are rarely reported, and point again to an algorithm with an embarassingly parallel classification.  On the other hand, the lack of scalabilty of the reading drags the overall performance down after a few hundred CPUs, and at the maximum CPU count is only about 65 \% efficient.  

The PFA algorithm has been shown to be highly optimized in terms of speedup and efficiency, yet one must wonder how it compares to other parallel software packages available for biological simulation analysis.  To the author's knowledge, the only other parallel MD analysis program is the CHARMM molecular dynamics package.  However, CHARMM is not optimized to run in a high performance environment of thousands of CPUs and therefore the comparison between PFA and CHARMM is not adequate.  To date, two of the most scalable and canonical simulation packages, NAMD and AMBER, could be modified to incorporate the calculations already built into the PFA program.  However, these packages employ a variety of computational tricks including smooth particle mesh Ewald\cite{pme95} and multipole methods\cite{namd99} for long range electrostatics, as well as spacial and particle-based parallel decompositions.  This translates the system being essentially split amongst compute nodes, and in order to add analysis functionality, extra communication per simulation step would be necessary.  This cost would inevitably lead to decreased performance, and one can imagine that keeping the analysis and simulation separate would lead to the most efficient use of computational resources. That said, a comparison will be made assuming the best possible scenario from AMBER and NAMD in order to compare the overall scalability of PFA.

\begin{figure}
  \centering
  \includegraphics*[width=10cm]{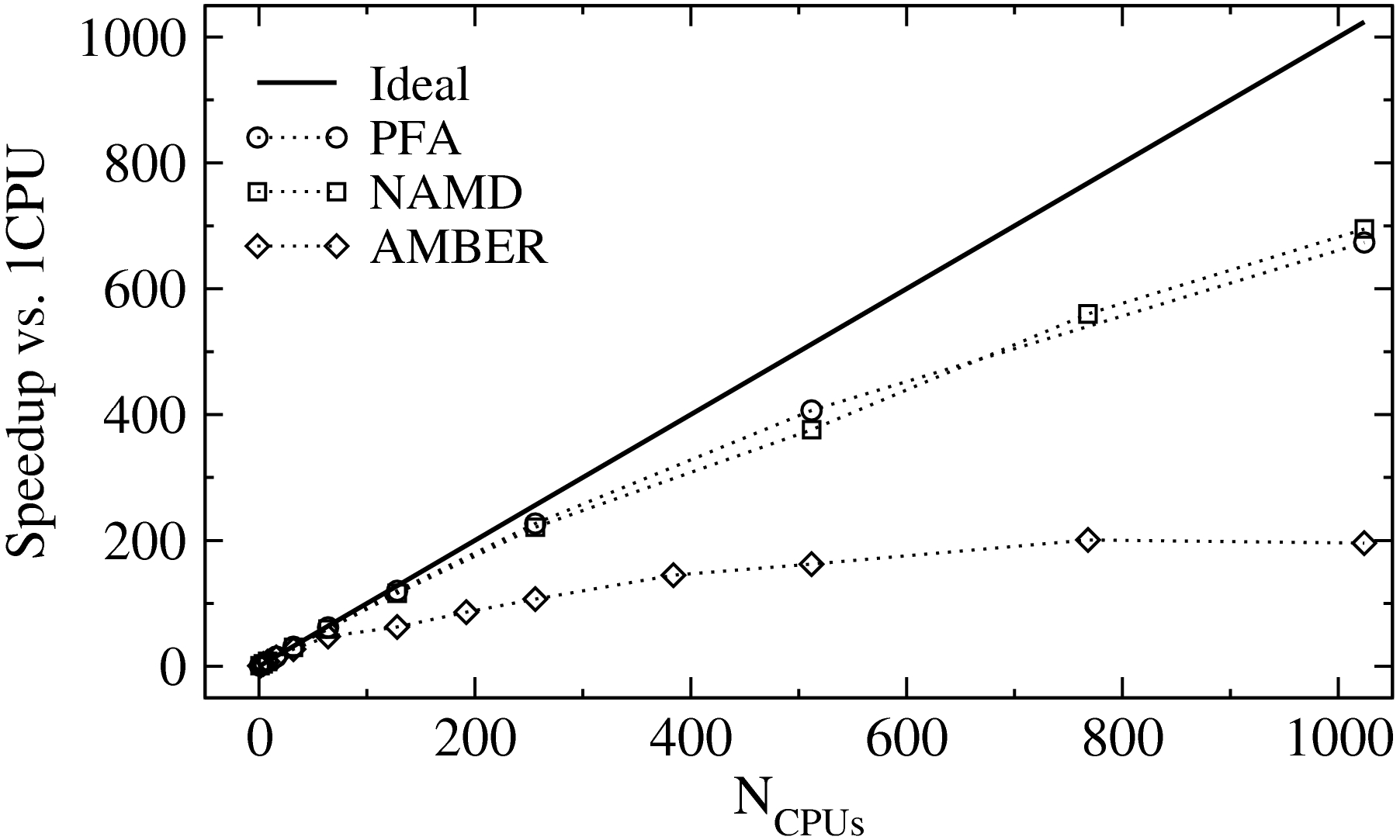}
  \caption{Comparison of the parallel scalability of the analysis code against two leading MD software packages, NAMD and AMBER.  The NAMD performance is measured for the Apo-1 protein, AMBER performance is for Cellulose, while PFA is tested using the BRC aqueous micelle.}
  \label{fig:speedup-compare}
\end{figure}

Shown in Fig. \ref{fig:speedup-compare} are the results for the total PFA program (circles), along with the best scalability results for both NAMD (squares) \cite{namdScale} and AMBER (diamonds)\cite{amber9scale} on modern supercomputers.  As can be seen, all three scale ideally to roughly 100 processors.  However after about 300 CPUs all three programs lose scalability, with AMBER degrading the fastest with more CPUs.  At the highest CPU count, AMBER is more than 3x less scalable than PFA, while PFA and NAMD are nearly equivalent.  It should be noted that at 512 CPUs, PFA outperforms both NAMD and AMBER in terms of efficiency and speedup.  Even more impressive, this comparison reports the performance metrics for AMBER and NAMD which are their respective best, while PFA used just a small trajectory with a rather small number of calculations per analysis run.  This means that Gustafson's law can be invoked even further to provide increased scalability by increasing the calculations per analysis run, or even more simply by increasing the trajectory length.

\section{Conclusions}
\paragraph*{}
Overall, the PFA analysis algorithm has been shown to be embarrassingly parallel over thousands of CPUs on the Ranger Supercomputer.  Due to the poor performance and efficiency of parallel reading, the total speedup of the PFA software degrades after a few hundred CPUs and is only about 65 \% efficent at 1,024 compute nodes.  In fact, the PFA software outperforms two MD software packages with similar analysis functionality at $N_{CPUs} > 300$.  The software, already involved in more than 10 CPU-years of production data analysis, will become more useful in time as more functions are added to the growing library, and as larger biological systems and long simulations times force the use of parallel algorithms into everyday life.  One can envision the utility of this software in analysis of replica exchange simulations, simulations of virus capsids, as well as mixed biomolecular systems as found in membrane bound protein simulations.

\section{Acknowledgments}
I am forever indebted to my advisor, Dmitry Matyushov, whose years of enlightening discussions and financial support made this code possible.  A special thanks goes to David Case and the Amber Development Team for granting permission of the use of their toplogy file reader code.  Also, I would like to recognize Vitaliy Kapko who suggested the current data partitioning scheme, and for helping to fish out many mistakes with the energy gap calculations.  D.N.L was supported by the NSF (CHE-0616646), and computer time at Ranger was provided by the TeraGrid's DAC program (TG-MCB080080N).


\bibliographystyle{cpc}
\bibliography{preprint}

\begin{thebibliography}{10}

\bibitem{namd99}
Schlick, T. et~al.,
\newblock Journal of Computational Physics {\bf 151} (1999) 9.

\bibitem{amber8}
Case, D.~A. et~al.,
\newblock Journal of Computational Chemistry {\bf 26} (2005) 1668.

\bibitem{charmm83}
Brooks, B. et~al.,
\newblock Journal of Computational Chemistry {\bf 4} (1983) 187.

\bibitem{pme95}
Essmann, U. et~al.,
\newblock The Journal of Chemical Physics {\bf 103} (1995) 8577.

\bibitem{pcjpcb08}
LeBard, D. and Matyushov, D.,
\newblock Journal of Physical Chemistry B {\bf 112} (2008) 5218.

\bibitem{rcjpcb08}
LeBard, D.~N., Kapko, V., and Matyushov, D.~V.,
\newblock Journal of Physical Chemistry B  (2008).

\bibitem{matyushov07}
Matyushov, D.,
\newblock Accounts of Chemical Research {\bf 40} (2007) 294.

\bibitem{pcjcp08}
LeBard, D.~N. and Matyushov, D.~V.,
\newblock The Journal of Chemical Physics {\bf 128} (2008) 155106.

\bibitem{small03}
Small, D., Matyushov, D., and Voth, G.,
\newblock Journal of the American Chemical Society {\bf 125} (2003) 7470.

\bibitem{chaudhury07}
Chaudhury, S. and Cherayil, B.~J.,
\newblock The Journal of Chemical Physics {\bf 127} (2007) 145103.

\bibitem{gustafson88}
Gustafson, J.~L.,
\newblock Communications of the ACM {\bf 31} (1988) 532.

\bibitem{tip3p83}
Jorgensen, W.~L., Chandrasekhar, J., Madura, J.~D., Impey, R.~W., and Klein,
  M.~L.,
\newblock The Journal of Chemical Physics {\bf 79} (1983) 926.

\bibitem{namdScale}
Brunner, R.~K., Phillips, J.~C., and Kale, L.~V.,
\newblock Scalable molecular dynamics for large biomolecular systems,
\newblock in {\em Proceedings of the IEEE/ACM SC2000 Conference}, page
  Technical Paper 271, IEEE Press, 2000.

\bibitem{amber9scale}
Walker, R.,
\newblock A guide to running amber molecular dynamics simulations at san diego
  supercomputer center (sdsc),
\newblock \begin{verbatim}http://coffee.sdsc.edu/rcw/amber_sdsc\end{verbatim}.

\end{thebibliography}

\end{document}